\documentstyle[12pt]{article}
\def\beq{\begin{equation}}
\def\eeq{\end{equation}}
\def\bea{\begin{eqnarray}}
\def\eea{\end{eqnarray}}
\def\bq{\begin{quote}}
\def\eq{\end{quote}}

\def\CMP{{\it Commun.Math.Phys.} }

\def\SNC{{\it Suppl. Nuovo Cimento} }

\def\NC{{\it Nuovo Cimento} }

\def\PL{{\it Phys.Lett.} }
\def\PR{{\it Phys.Rev.} }
\def\PRL{{\it Phys.Rev.Lett.} }

\def\PTP{{\it Progr.Theor.Phys.} }

\def\gappeq{\mathrel{\rlap {\raise.5ex\hbox{$>$}}
{\lower.5ex\hbox{$\sim$}}}}

\def\lappeq{\mathrel{\rlap{\raise.5ex\hbox{$<$}}
{\lower.5ex\hbox{$\sim$}}}}

\begin{document}
\pagestyle{empty}
\begin{flushright}
{CERN-TH/2000-138}\\
{LAPTH-794/2000}
\end{flushright}
\vspace*{5mm}
\begin{center}
{\bf HARRY LEHMANN AND THE ANALYTICITY UNITARITY PROGRAMME}\footnote{To appear in
a volume of Communications in Mathematical Physics, dedicated to the memory of
Harry Lehmann.}
\\
\vspace*{1cm} 
{\bf Andr\'e MARTIN}\\
\vspace{0.3cm}
Theoretical Physics Division, CERN \\
CH - 1211 Geneva 23 \\
and\\
LAPP\footnote{URA 1436 du CNRS, associ\'ee \`a l'Universit\'e de Savoie.}\\
F - 74941  Annecy le Vieux Cedex \\
\vspace*{2cm}  
{\bf ABSTRACT} \\ \end{center}
\vspace*{5mm}
I try to describe the extremely fruitful interaction I had with Harry Lehmann and
the results which came out of the analyticity unitarity programme, especially the
proof of the Froissart bound, which, with recent and future measurements of total
cross-sections and real parts, remains topical.
 
\vspace*{2cm}
\begin{center}
Dedication\\
I dedicate this paper to Marie-No\"elle Fontaine, \\
the last of the many papers she typed so skillfully\\
for me, wishing her a happy retirement.
\end{center}
 
\vspace*{1cm}
\begin{flushleft} 
CERN-TH/2000-138 \\
LAPTH-794/2000\\
May 2000
\end{flushleft}
\vfill\eject

\setcounter{page}{1}
\pagestyle{plain}

My first meeting with Harry Lehmann was not with his person but with the famous
paper of the trio Lehmann-Symanzik-Zimmermann, LSZ \cite{aaa}, the importance of
which everybody in the Theory group of Maurice L\'evy at Ecole Normale realized
immediately. In spite of the fact that I did not know German (I still don't) I
read it, Nuovo Cimento in one hand, dictionary in the other hand (I am a
``corrected" left hander). Later Harry visited the Ecole Normale in person and I
was immediately impressed. That was the time where there was a wave of interest
into what is an unstable particle and Lehmann and L\'evy were some of the people
involved. I remember also quite vividly our meeting at the La Jolla Conference in
1961 which I attended, coming from CERN. It was, as I realized a posteriori, a
very important conference, for physicists and for people (some of the people I
met there became my very best friends). I remember that Marcel Froissart gave a
talk on his famous Froissart bound \cite{bb} on the total cross-section,
$\sigma_t < c \log s)^2$, $s$ square of the centre-of-mass energy, and Harry with
his very meticulous mind found out that some of the estimates of Froissart were
not quite correct, though this did not affect the result (some year later, I
published a sum rule on pion-nucleon scattering and Harry discovered a very well
hidden mistake. I was very impressed). Anyway we were both admirative of the
achievement of  Froissart and for me it was a decisive turning point, since I
left almost completely for many years potentials and the Schr\"odinger equation
for the study of high-energy scattering and high-energy bounds. The Froissart
bound was derived from a combination of  the Mandelstam representation \cite{cc}
where the scattering amplitude is the boundary value of an analytic function of
two variables, which implies automatically dispersion relations proved from field
theory \cite{dd} in one variable as well as the Lehmann ellipse \cite{ee} which
is probably the most celebrated result of Harry, a fundamental result  presented
in 10 small pages of Nuovo Cimento (compare with the incredibly lengthy papers on
what I would call ``rigorous atomic physics" which appeared during the last 15
years!).

The trouble with the Mandelstam representation is that nobody was ever able to
prove it even in perturbation theory (through some wrong proofs were published!).
Both Harry and I were anxious to obtain high-energy bounds with minimal
assumptions. A step in this direction was made by Greenberg and Low \cite{ff} who
used the Lehmann Ellipse to derive a bound on the total cross-section where
$(\log s)^2$ was replaced by $s(\log s)^2$. Myself, I realized that the whole
Mandelstam representation was not needed to get the Froissart bound and that it
was sufficient to replace the Lehmann ellipse by a larger one \cite{ggg}. Later,
in Princeton, Y.S. Jin (a former student of Harry) and I found a way to control
the growth of the scattering  amplitude for unphysical momentum transfer using
positivity \cite{hh} but at the time we made no progress on the derivation of
the Froissart bound. In the autumn of 1965 I was visiting IHES (Institut des
Hautes Etudes Scientifiques) and Harry was there. He attracted my attention on a
paper by Nakanishi which contained the claim that the Lehmann Ellipse could be
enlarged by using results from perturbative field theory, leading to the
obtention of the Froissart bound. As I shall explain later, we tried to make
sense of the paper of Nakanishi \cite{jj} but in the end could not. Nevertheless
it started again my interest in the subject, and after a visit to Cambridge where
I learnt that the Nakanishi perturbative domain of analyticity \cite{kk} had been
obtained independently and in a simpler way by T.T. Wu \cite{lll}, I came back
to CERN and finally succeeded, using positivity properties not terribly different
from those I had used with Jin, to enlarge the Ellipse without using perturbation
and prove the Froissart bound from first principles. Something rather rare
happened: Harry sent a postcard to congratulate me, but while moving from one
apartment to another one or maybe from one office to another, I lost it!

I had many occasions to meet Harry later, but the last one was in the spring of
1998 at CERN where he came to work with T.T. Wu after an operation which seemed
successful. At the Ringberg Castle meeting, in the honor of Wolfhart
Zimmermann, he was supposed to be the first speaker and could not come because
he was ill. I became the first speaker. Then I knew I would never meet him
again.

Now I believe that it is necessary to give some technical details.

In 3+1 dimensions (3 space, 1 time) the scattering amplitude depends on two
variables energy and angle. For a reaction $A + B \rightarrow A + B$
\beq
E_{c.m.} = \sqrt{M^2_A+k^2} + \sqrt{M^2_B+k^2}~,
\label{one}
\eeq
$k$ being the centre-of-mass momentum. The angle is designated by $\theta$.
There are alternative variables:
\beq
s = (E_{CM})^2~,~~~t = 2k^2 (\cos\theta -1)
\label{two}
\eeq(Notice that physical $t$ is NEGATIVE).

We shall need later an auxiliary variable $u$, defined by 
\beq
s + t + u = 2 M_A^2 + 2M^2_B
\label{three}
\eeq

The \underline{Scattering amplitude} (scalar case) can be written as a partial
wave expansion, the convergence of which will be justified in a moment:
\beq
F (s,\cos\theta ) = {\sqrt{s}\over k} \sum (2\ell +1) f_\ell(s) P_\ell
(\cos\theta)
\label{four}
\eeq
$f_\ell(s)$ is a partial wave amplitude.

The \underline{Absorptive part}, which coincides for $\cos\theta$ real 
(i.e., physical) with the imaginary part of $F$, is defined as
\beq
A_s (s,\cos\theta ) = {\sqrt{s}\over k} \sum (2\ell + 1) ~{\rm Im}~ f_\ell (s)
(\cos\theta)
\label{five}
\eeq

The \underline{Unitarity condition}, implies, with the normalization we have
chosen
\beq
{\rm Im}~ f_\ell (s) \geq  \vert f_\ell (s) \vert^2
\label{six}
\eeq
which has, as a consequence
\beq
{\rm Im}~ f_\ell (s) > 0~,~~~ \vert f_\ell  \vert < 1~.
\label{seven}
\eeq
The differential cross-section is given by
$$
{d\sigma\over d\Omega} = {1\over s}~~\vert F\vert^2~,
$$
and the total cross-section is given by the ``optical theorem"
\beq
\sigma_{total} = {4\pi\over k\sqrt{s}}~~A_s(s,\cos\theta = 1)~.
\label{eight}
\eeq
With these definitions, a dispersion relation can be written as:
\beq
F(s,t,u) = {1\over\pi} \int{A_s(s^\prime,t)ds^\prime\over s^\prime -s} +
{1\over\pi}
\int {A_u(u^\prime,t)du^\prime\over u^\prime -u}
\label{nine}
\eeq
with possible subractions, i.e., for instance the replacement of $1/(s^\prime
- s)$ by
$s^N /s^{\prime N}(s^\prime - s)$
and the addition of a polynomial in $s$, with coefficients depending on $t$. 

The scattering amplitude in the $s$ channel $A+B\rightarrow A+B$ is the
boundary value of $F$ for $s + i\epsilon$, $\epsilon > 0  \rightarrow 0$, $s >
(M_A+M_B)^2$. In the same way the amplitude for $A+\bar B\rightarrow A+\bar B$, 
$\bar B$ being the antiparticle of $B$ is given by the boundary value of $F$
for $u+i\epsilon , ~~\epsilon\rightarrow 0 ~~u > (M_A+M_B)^2$. Here we
understand the need for the auxiliary variable $u$.

The dispersion relation implies that, for fixed $t$ the scattering amplitude
can be continued in the $s$ complex plane with two cuts. The scattering
amplitude possesses the reality property, i.e., for $t$ real it is real between
the cuts and takes complex conjugate values above and below the cuts.

In the most favourable cases, like $\pi\pi\rightarrow\pi\pi$ or $\pi N\rightarrow
\pi N$ scattering
 dispersion relations have been established for
$-T < t \leq 0$,  $T > 0$ \cite{dd}.

In the general case, even if dispersion relations are not proved, the crossing
property of Bros, Epstein and Glaser states that the scattering amplitude is
analytic in a twice cut plane, minus a finite region, for \underline{any
negative} $t$
\cite{mm}. So it is possible to continue the amplitude directly from
$A+B\rightarrow A+B$ to the complex conjugate of $A+\bar B\rightarrow A+\bar
B$. By a more subtle argument, using a path with fixed $u$ and fixed $s$ it is
possible to continue directly from
$A+ B\rightarrow A+ B$ to $A+\bar B\rightarrow A+\bar B$

At this point, we see already that one cannot dissociate analyticity, i.e.,
dispersion relations, and unitarity, since the discontinuity in the dispersion
relations is given by the absorptive part. In the simple case of $t = 0$, the
absorptive part is given by the total cross-section and the forward amplitude 
is given, as we said already for the case of Compton Scattering, by an
integral over physical quantities.

It was recognized very early that the combination of analyticity and unitarity
might lead to very interesting consequences and might give some hope to fulfill
at least partially the $S$ matrix Heisenberg program. This was very clearly
stated already in 1956 by Murray Gell-Mann \cite{nn} at the Rochester
conference. Later this idea was taken over by many people, in particular by
Geff Chew. To make this program as successful as possible it seemed necessary
to have an analyticity domain
as large as possible. Dispersion relations are fixed $t$ analyticity
properties, in the other variable $s$, or $u$ as one likes.

Another property derived from local field theory was the existence of the
\underline{Lehmann ellipse} \cite{ee}, which states that for fixed $s$,
physical, the scattering amplitude is analytic in $\cos\theta$ in an ellipse
with foci at $\cos\theta = \pm 1$. $\cos \theta = 1$ corresponds to $t = 0$ the
ellipse therefore contains a circle 
\beq
\vert t \vert < T_1(s)~.
\label{ten}
\eeq
$T_1(s)$ is given by
$$
\matrix{
x_0 &= 1 + {T_1(s)\over 2k^2} \hfill\cr
&\cr
x_0 &= \left[ 1 + {(M^2_1-M^2_A)~(M^2_2-M^2_B)\over
k^2(s-(M_1-M_2)^2)}\right]}^{1/2}
$$
where $M_A$ and $M_B$ are the masses of the particles, $M_1$ and $M_2$ are the
lowest intermediate states in the currents associated to the fields of the
incoming particles.

Hence $T_1(s)\rightarrow 0$ for $s\rightarrow (M_A+M_B)^2$
and $s\rightarrow \infty$. 

The absorptive part is analytic in the larger ellipse, the ``large" Lehmann
ellipse, containing the circle 
\beq
\vert t \vert < T_2(s)~,
\label{eleven}
\eeq
$T_2$ is given by
$$
2x_0^2 - 1 = 1 + {T_2(s)\over 2k^2}
$$
So $T_2(s)\rightarrow c > 0$ for $s \rightarrow (M_A+M_B)^2$, $T_2(s)
\rightarrow 0$ for $s\rightarrow\infty$.

It was thought by Mandelstam that these two analyticity properties, dispersion
relations and Lehmann ellipses, were insufficient to carry very far the
analyticity-unitarity program. he proposed the Mandelstam representation
\cite{cc} which can be written schematically as

\bea
F = &&{1\over \pi^2} \int {\rho (s^\prime, t^\prime)ds^\prime dt^\prime \over
(s^\prime -s)~(t^\prime -t)}\nonumber \\
&&+ {\rm circular~permutations~in}~ s, t, u \nonumber \\
&&+ {\rm one~dimensional~dispersion~integrals} \nonumber \\
&&+ {\rm subtractions}
\label{twelve}
\eea

This representation is nice. It gives back the ordinary dispersion relations
and the Lehmann ellipse when one variable is fixed, but it was never proved nor
disproved for all mass cases, even in perturbation theory,. One contributor,
Jean Lascoux, refused to co-sign a ``proof", which, in the end, turned out to be
imperfect.

One very impressive consequence of Mandelstam representation was the proof, by
Marcel Froissart, that the total cross-section cannot increase faster than (log
$s)^2$, the so-called ``Froissart Bound" \cite{bb}.

My own way to obtain the Froissart bound \cite{ggg} was to use the fact that the
Mandelstam representation implies the existence of an ellipse of analyticity in
$\cos\theta$ qualitatively \underline{larger} than the Lehmann ellipse, i.e.,
such that it contains a circle $\vert t\vert < R$, $R$ fixed, independent of
the energy. This has a consequence that ${\rm Im}~ f_\ell(s)$ decreases with
$\ell$ at a certain exponential rate because of the convergence of the Legendre
polynomial expansion and of the polynomial boundedness, but on the other hand
the ${\rm Im}~ f_\ell(s)$'s are bounded by unity because of unitarity [Eq.
(\ref{seven})]. Taking the best bound for each $\ell$ gives the Froissart bound.

Let me now try to recall the exchange Harry Lehmann and I had in the Autumn of
1965 in Bures sur Yvette. We had in common the same desire to find a proof of the
Froissart bound without using the Mandelstam representation  and to find a way to
enlarge the Lehmann ellipse. Harry pointed out to me a paper published by N.
Nakanishi \cite{jj} a few months earlier where he claimed that he had a proof of
the Froissart bound.

	Let me remind you that the largest possible ellipse of convergence of the
Legendre Polynomial series for the  absorptive part has necessarily a singularity
at its right extremity. This is the analogue of a classical theorem on power
series with positive coefficients. This means that if you succeed (take the
$\pi\pi$ case, $m_\pi = 1$) in proving that the absorptive part is analytic in the
neighbourhood of the segment
$$
t = 0 \quad\quad t = 4~,
$$
then it is automatically analytic in the ellipse with foci
$$
t = 4-s \quad\quad t = 0~,
$$
and right extremity $t = 4$, and a fortiori it is analytic in the circle
$$
\vert t\vert < 4~,
$$
entirely contained in the ellipse.

	Nakanishi had obtained a representation valid for any Feynman diagram \cite{kk}
$$
T_N(s,t) = \int {d^n \alpha \over \left[ f(\alpha ) + s ~g(\alpha ) +
t~h(\alpha)\right]^p}
$$
Later on I learnt from P. Landshoff that this representation had also been
obtained, independently and in a simpler way by T.T. Wu \cite{lll}. A minimal
analytic domain for $T_N(s,t)$ is obtained when the denominator in the integral
representaiton does not vanish.

This domain, for the $\pi\pi$ case  for fixed complex $s$ is a kind
of strip containing the straight line going through $t = 0$ and $t = 4-s$, (which
corresponds to $\cos\theta_s$ real), and the segments $-4 < t < +4$ and $-s < t <
8-s$.
When $s$ tends to a real value the domain shrinks to zero for $s > 4$ and for 
$s < -t$ (for $t$ real). 

This means that for $t$ fixed, real $-4 < t < +4$, dispersion relations
hold.

The Nakanishi-Wu representation also implies the validity of partial wave
dispersion relations but this is irrelevant for our problem.

However, there is nothing like a small or a large Lehmann ellipse in this domain.
The absorptive part in perturbation theory, which is defined only in the limit $s
\rightarrow s_R + i\epsilon$, $\epsilon\rightarrow 0$ has a priori no analyticity
in $t$. A priori, it is just a distribution.

In fact in perturbation theory, unitarity connects amplitudes of different orders
and positivity properties of the absorptive part are completely hidden. In
three-space dimensions, nobody knows if the perturbation series can be resummed
(probably not!) and it is not ``legal" to combine the results of axiomatic field
theory and perturbation theory. Of course one can always try it as a game, which
is what Harry and I tried to do, but we went nowhere.

It is only in December 1965, after a visit to Cambridge, that I found a way to
enlarge the Lehmann ellipse in the framework of axiomatic field theory \cite{oo},
without using at all the results of perturbation theory. I was maybe a bit unfair
not to quote the Nakanishi-Wu representation because the ``wrong" paper of
Nakanishi was undoubtedly a source of stimulation but, on the other hand, I did
not use it at all.

	Our method was the following.

The positivity of ${\rm Im}~ f_\ell$ implies, by using expansion (\ref{five}),
\beq
\left\vert \left({d\over dt}\right)^n A_S(s,t)\right\vert
_{-4k^2 \leq t \leq 0}
 \leq
\left\vert\left({d\over dt}\right)^n A_S (s,t) \right\vert_{t=0}~.
\label{thirteen}
\eeq
To calculate 
$$F(s,t) = {1\over\pi}~~\int_{s_0} {A_s(s^\prime t) ds^\prime \over s^\prime -s}
$$
(forget the left-hand cut and subtractions!), for $s$ real $< s_0$ one can
expand $F(s,t)$ around $t = 0$. From the property (\ref{thirteen}) one can
prove that the successive derivatives can be obtained by differentiating
\underline{under the integral}. When one resums the series one discovers that
this can be done not only for $s$ real $<s_0$, but for any $s$ and that the
expansion has a domain of convergence in $t$ \underline {independent of $s$}.
This means that the large Lehmann ellipse \underline{must} contain a circle
$\vert t \vert < R$. This is exactly what is needed to get the Froissart bound.
In fact, in favourable cases, $R = 4m^2_\pi$, $m_\pi$ being the pion mass. A
recipe to get a lower bound for $R$ was found by Sommer \cite{pp}

\beq
R \leq {\rm sup}_{s_0 < s <\infty} T_1(s)
\label{forteen}
\eeq 
It was already
known that for
$\vert t
\vert < 4m^2_\pi$ the number of subtractions in the dispersion relations was at
most
\underline{two}
\cite{hh}, and it lead to the more accurate bound \cite{qq}
\beq
\sigma_T < {\pi\over m^2_\pi}~~(\log s)^2
\label{fifteen}
\eeq

Notice that this is \underline{only} a \underline{bound}, not an asymptotic
estimate.

In spite of many efforts the Froissart bound was never qualitatively improved,
and it was shown by Kupsch \cite{rr} that if one uses only ${\rm Im}~ f_\ell
\geq
\vert f_\ell\vert^2$ and full crossing symmetry one cannot do better than
Froissart. 

On the more theoretical side one might wonder if using crossing symmetry and
analytic completion one could not prove Mandelstam representation at least for
the pion-pion case using only axiomatic results. This is not the case, as I
showed it in 1967 at a meeting organized by Bob Marshak in Rochester where Harry
was present
\cite{ss}. One can write a representation of the scattering amplitude
$$
\matrix{
F_\nu = &\int^1_0~~dx~\int^\infty_{p_0q_0}~
{dp~dq~w(x,p,q)\over x(p-s)^\nu + (1-x)~(q-t)^\nu} \cr
&\cr
& + {\rm circular~permutations}}
$$ 

For $\nu = 1/2$ this is just a funny way to write the Mandelstam representation.

For $\nu = 1$, you get back to the Nakanishi-Wu representation.

For $\nu = 2/3$ you get a natural domain bigger than all you can get from 
axiomatic field theory and positivity.

Before 1972, rising cross-sections were a pure curiosity. Almost everybody
believed that the proton-proton cross-section was approaching 40 millibarns at
infinite energy.  Yet, Khuri and Kinoshita \cite{tt} took seriously very early the
possibility that cross-sections rise and proved, in particular, that if the
scattering amplitude is dominantly crossing even, and if $\sigma_t \sim (\log
s)^2$ then
$$
\rho = {Re F\over Im F} \sim {\pi\over \log s}~,
$$
where $Re F$ and $Im F$ are the real and imaginary part of the forward
scattering amplitude.

As early as 1970, Cheng and Wu proposed a model in which cross-sections were
rising \cite{uu} and eventually saturating the Froissart bound. However, at that
time there was no experimental indication of this. It is only in 1972 that it was
discovered at the  ISR, at CERN, that the $p-p$ cross-section was rising by 3
millibarns from 30 GeV c.m. energy to 60 GeV c.m. energy
\cite{vv}. I suggested to the experimentalists that they should measure $\rho$
and test the Khuri-Kinoshita predictions. They did it \cite{ww} and this kind
of combined measurements of $\sigma_T$ and $Re F$ are still going on. In
$\sigma_T$ we have now more than a 50 \% increase with respect to low energy
values. For an up to date review I refer to the article of Matthiae
\cite{yy}. it is my \underline{strong conviction} that this activity should be
continued with the future LHC. A breakdown of dispersion relation might be a
sign of new physics due to the presence of extra compact dimensions of space
according to N.N. Khuri \cite{zz}. Future experiments, especially for $\rho$,
will be difficult because of the necessity to go to very small angles, but not
impossible \cite{aai}.
\vfill\eject

\end{document}